\documentclass[%
 reprint,
superscriptaddress,
 amsmath,amssymb,
 aps,
]{revtex4-1}

\usepackage{graphicx}
\usepackage{siunitx}

\begin{document}
\title{Shipping traffic through the Arctic Ocean: spatial distribution, temporal evolution and its dependence on the sea ice extent}

\author{Jorge P. Rodríguez}
    \email{jorgeprodriguezg@gmail.com}
    \affiliation{Instituto de Física Interdisciplinar y Sistemas Complejos (IFISC), CSIC-UIB, 07122 Palma de Mallorca (Spain)}
    \affiliation{CA UNED Illes Balears, 07009 Palma (Spain)}
    \affiliation{Instituto Mediterráneo de Estudios Avanzados (IMEDEA), CSIC-UIB, 07190 Esporles (Spain)}
\author{Konstantin Klemm}
    \affiliation{Instituto de Física Interdisciplinar y Sistemas Complejos (IFISC), CSIC-UIB, 07122 Palma de Mallorca (Spain)}
\author{Carlos M. Duarte}
    \affiliation{Red Sea Research Center (RSRC), King Abdullah University of Science and Technology (KAUST), 23955 204 Thuwal (Saudi Arabia)}
\author{Víctor M. Eguíluz}
    \affiliation{Basque Centre for Climate Change (BC3) (Spain)}
    \affiliation{IKERBASQUE, Basque Foundation for Science (Spain)}

\begin{abstract}
 \textbf{Background}
 
 The reduction in sea ice cover with Arctic warming facilitates the transit of ships through routes that are remarkably shorter than the traditional shipping routes. Automatic Identification System (AIS), ideally designed to avoid vessel collisions, transmits on vessel navigation information (currently 27 types of messages) such as name, position or speed, is a powerful data source to monitor the progress of Arctic shipping as the ice cover decreases.
 
 \textbf{Results}
 
 Based on the analysis of an online platform collecting shipping AIS data, we quantified the spatial distribution of shipping through the Arctic Ocean, its intensity and the temporal evolution, in relation to the area released by the sea ice area. Shipping through the Arctic Ocean is distributed spatially following a heavy-tailed distribution, implying heavy traffic through a limited Arctic area, with an exponent that depends on the vessel category. 
 
 Fishing is the category with the largest spatial spread, with the width of shipping routes correlated with the proximal sea ice area. The time evolution of these routes is characterized by increasing extended periods of shipping activity through the year.
 
 \textbf{Conclusions}
 
AIS data offers valuable information on the activity of the international fleet worldwide. In the context of the new international agreements, it is a valuable source to monitor shipping, fishing and the potential impact in marine life among other aspects. Here we have focused on the Arctic shipping in recent years, which is rapidly growing, particularly around the Northeastern and Northwest Passage coastal routes, providing an opportunity for the design of shorter shipping routes and reduced greenhouse gas emissions from transport of goods, but at a risk of impacts on the Arctic ecosystem.
\end{abstract}

\maketitle
\section{Background}
\label{sec:background}

Shipping traffic represents the dominant transportation mode in global trade, delivering more than 80\% of the volume of the international trade of goods \cite{unctad2021rev}. In fact, economic growth has led to a parallel increase in marine traffic 60\% in the period between 1992 and 2002 \cite{tournadre2014anthropogenic}. The importance of maritime transport to the global economy was evidenced in 2020 when the Suez Canal was blocked when the Ever Given container ship got stranded. The opportunity to leverage the opening of new Arctic routes delivering goods from Asia to Europe and North America due to the decrease of the ice cover will increase traffic and bring new threats to this vulnerable ecosystem \cite{Melia2016}, adding to the direct impacts of rapid climate change in the Arctic. In fact, an estimation of the shipping time has revealed an increase in 7\% between 2013 and 2022 \cite{muller2023arctic}.

Tracking technologies are playing a major role in the analysis of vessels' movement through the oceans, allowing the quantification of multiple vessel behaviors with economic, political and ecological consequences. For example, tracking of fishing vessels facilitated the inference of hot spots of fishing activity \cite{kroodsma2018tracking,rodriguez2021global,frawley2022clustering}. Moreover, the overlap between fishing vessels' trajectories and movement tracking of marine animals has revealed the regions with a high risk of overlap and thus risk of bycatch between fishing vessels and sharks \cite{queiroz2019global}, and the collision risk of large vessels and whale sharks \cite{womersley2022global}. 

Currently, products derived from vessel tracking data are openly available. For example, Global Fishing Watch's main product describes fishing effort at high spatial and temporal resolution globally \cite{kroodsma2018tracking}. However, broader datasets, including for instance the trajectories of other vessel categories globally, are available under private purchase. To overcome the problem of data ownership and standardization, new initiatives are being developed to perform online analyses with access to previously clean, pre-processed and scientifically validated datasets from multiple sources. In this direction HUB Ocean has developed the Ocean Data Platform (oceandata.earth), where scientists can perform online analyses of multiple datasets describing diverse oceanic phenomena, such as parasite infections in fish farms, global vessel emissions or the geospatial data describing Marine Protected Areas (MPAs). 

Here, we report our analysis of the shipping traffic on the Arctic Ocean between January 2020 and April 2022 through a data analysis developed on a Private Preview Week of the Ocean Data Platform.

\section{Data Description}
\begin{figure*}[bt!]
\centering
\includegraphics[width=\textwidth]{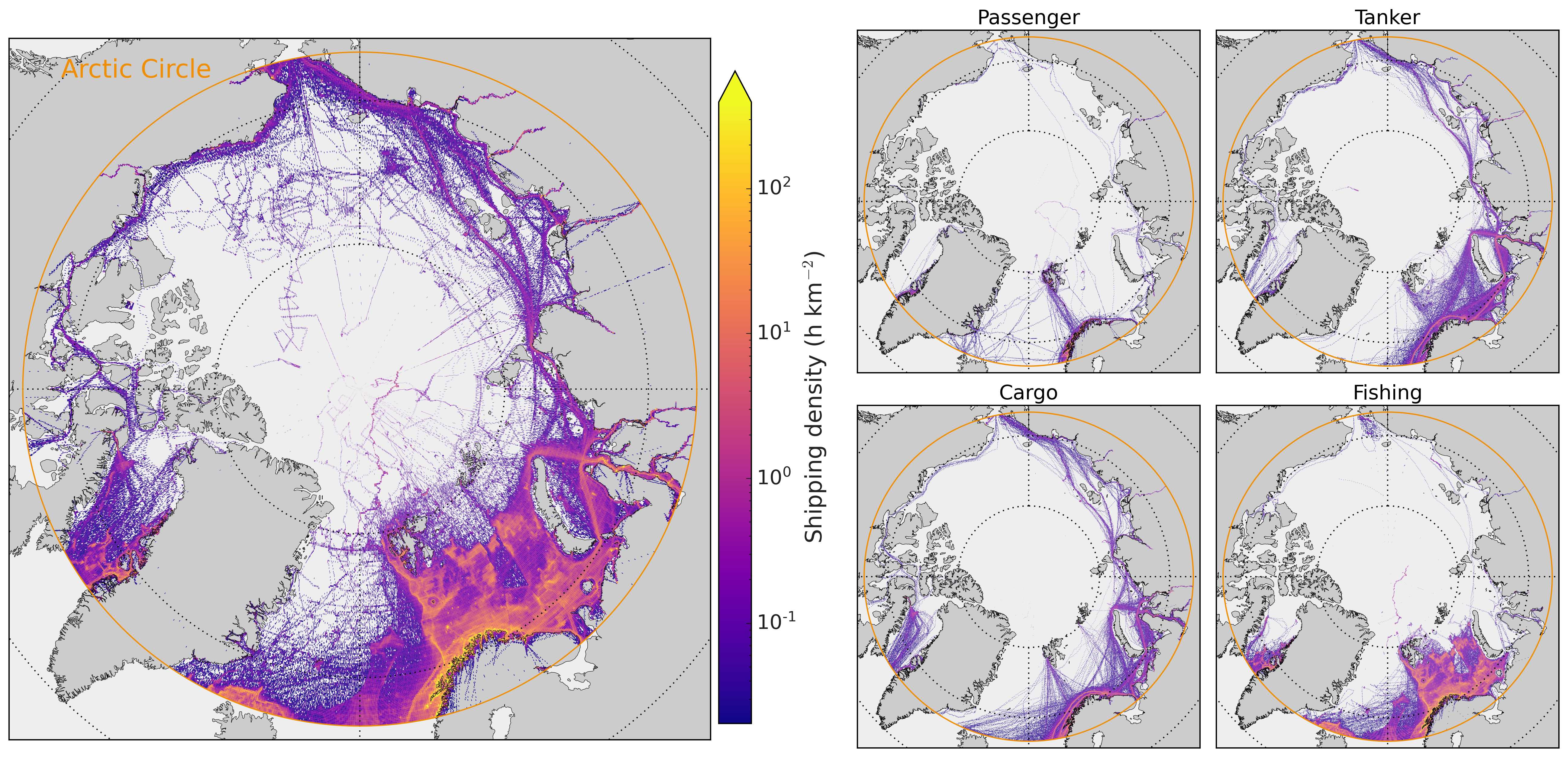}
\caption{Arctic shipping between January 2020 and April 2022. Shipping density is computed as the aggregated transit time over all the vessels, at each $0.1^{\circ}\times 0.1^{\circ}$ cell, divided by the cell area. Left panel represents the total shipping traffic, while right panels include the traffic for passenger, tanker, cargo and fishing vessel categories.}\label{fig:f1}
\end{figure*}

{\bf Shipping traffic.} The space use of ships transiting the Arctic Ocean was inferred from the Automatic Identification System (AIS) data. AIS is a system introduced for maritime safety that provides, among different data variables, the speed, latitude and longitude of the vessels using the system. The platform Ocean Data Platform aggregated AIS tracking data at a monthly resolution with a high spatial resolution and reported the number of hours each grid cell has been transited by vessels. The monthly transiting time was available from January 2020 to April 2022, and it specified five vessel categories (cargo, fishing, passenger, tanker and other). For our analysis, we introduced a global grid of $0.1^{\circ}\times 0.1^{\circ}$ resolution, selecting latitudes higher than the Arctic Circle ($66.6^{\circ}$, Fig. \ref{fig:f1}).

The time evolution of the area covered by the shipping routes was complemented with the dataset used for a previous assessment of the Arctic shipping traffic \cite{eguiluz2016quantitative}, to provide a comparison of the Ocean Data Platform dataset and illustrate the long-term time evolution of the shipping traffic in the Arctic Ocean. This previous dataset reported the monthly number of unique vessels detected by the AIS system in each $0.25^{\circ}\times 0.25^{\circ}$ grid cell, between July 2010 and May 2015.

{\bf Sea ice cover.} Sea ice area was obtained from the Sea Ice Index, provided by the National Snow and Ice Data Center (United States) \cite{seaice}. This dataset reported the monthly evolution of the sea ice area in the Northern Hemisphere, as well as specific subregions of the Arctic, where we considered the Canadian Archipelago Area, the Baffin Bay and the Beaufort Sea for the Northwest Passage route, while we considered the East Siberian Sea, the Kara Sea and the Barents Sea for the Northeastern route.

\section{Analyses}

The shipping density was computed as the transit time in each grid cell divided by the cell area, aggregating the time of all the ships using AIS in the considered region. This pattern revealed hot spots of shipping activity, both in the overall map and in the specific patterns associated with different vessel categories (Fig. \ref{fig:f1}). Fishing vessels represented the largest contribution to shipping in the Arctic Ocean, especially in the Barents Sea but also in the proximity of Iceland. Cargo vessels, similarly to tanker vessels, displayed patterns where we observed the Northeastern and Northwest Passage routes, with the latter becoming broader in the Baffin Bay, where tanker trajectories occupied less area in this region. Passenger traffic covered lower fractions of area as the most frequented routes were shorter, for example on the Norwegian and Icelandic coasts.   
\begin{figure*}[bt!]
\centering
\includegraphics[width=\textwidth]{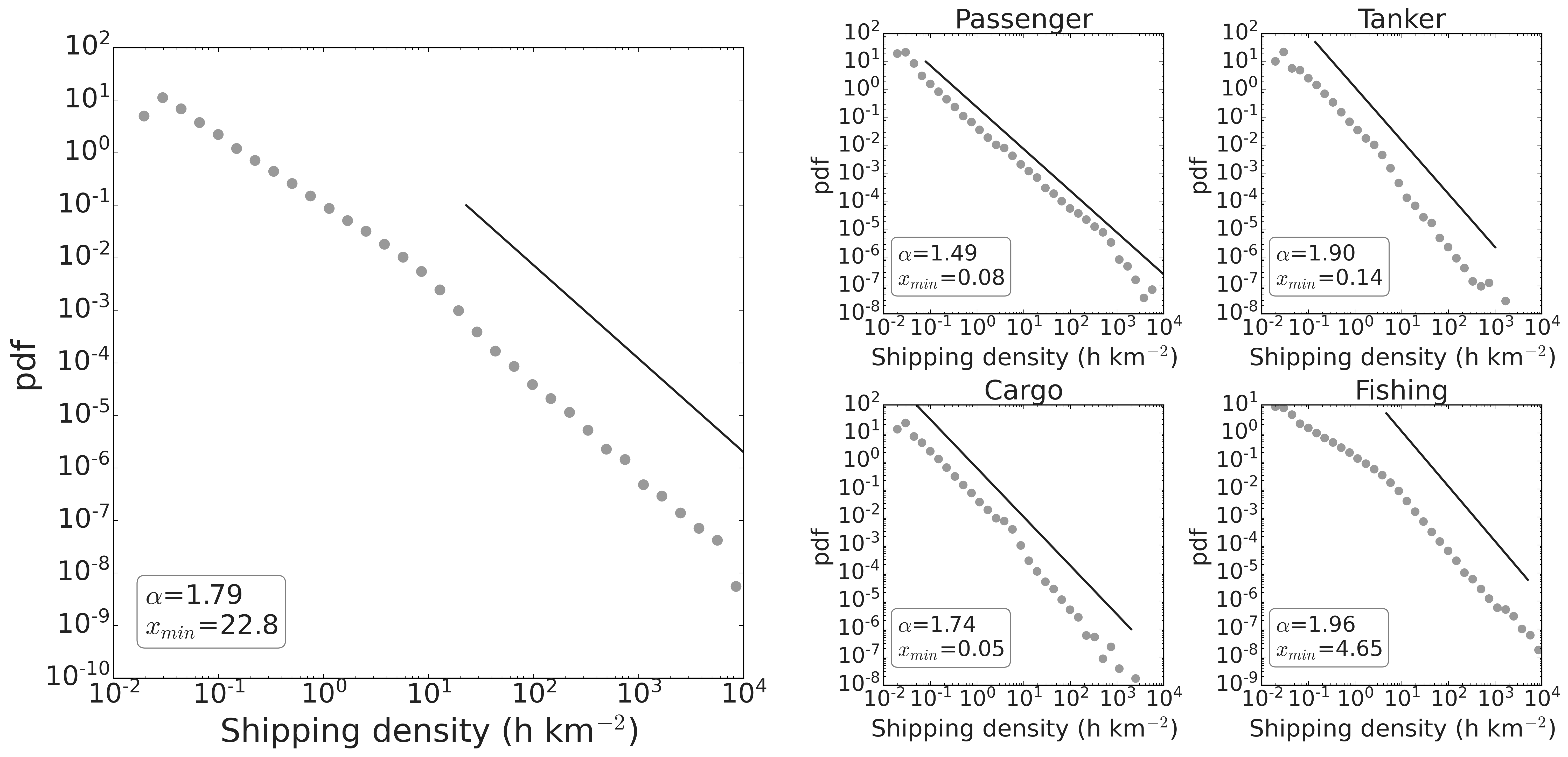}
\caption{Shipping density distributions for total shipping (left) and each vessel category (right). These heavy-tailed distributions fit to a power-law distribution $pdf(x)\sim x^{-\alpha}$ for $x>x_{min}$. The slope, obtained from fitting with the Python package \emph{powerlaw} \cite{alstott2014powerlaw}, is represented in the black lines }\label{fig:f2}
\end{figure*}

The shipping density heterogeneity across the space was described by heavy-tailed distributions, such that most grid cells displayed a low shipping density, with a few cells concentrating large values (Fig. \ref{fig:f2}). Specifically, the shipping density distributions for the aggregated (across categories) and for specific categories were described by power-law distributions. We performed a power-law regression to these distributions with the Python package \emph{powerlaw}, obtaining the fitted exponents 1.79 (aggregated), 1.49 (passenger), 1.90 (tanker), 1.74 (cargo) and 1.96 (fishing). For fishing, we observed two regimes, where the distribution is closer to a uniform distribution for low densities (i.e., with a smaller exponent), while large densities implied a faster decrease, which was the behaviour captured by the regression. 

We propose a null model of transit between geographical locations to understand the differences between those exponents, and we apply it to one- and two-dimensional systems. In a one-dimensional system, ships can move through space, modifying their coordinate $x$ and always with a constant and positive speed $v=\frac{\text{d}x}{\text{d}t}$. We consider a destination located at $x=1$, and the origins distributed uniformly in the interval $[0,1)$, with vessels departing randomly from any of these origins. We aim at computing the distribution of the transit times across different locations. The total transit time between $x$ and $x+\text{d}x$ will be given by the number of vessels that crossed that region, that is, $Nx$, with $N$ being the total number of vessels, and $x$ being the probability that a vessel departs from a location lower than $x$, times $\text{d}x/v$ (i.e., the transit time of one vessel), implying that the shipping density $\rho_{1d}$, i.e. the transit time per unit length, is
\begin{equation}
    \rho_{1d} (x)= \frac{Nx}{v}
    \label{eq:rho1d}
\end{equation}

Thus, the cumulative distribution function (CDF) of the shipping density $\rho_{1d}$ is the length of the segment with locations that have a shipping density time lower than $\rho_{1d}$. This length is $x$ in Eq. (\ref{eq:rho1d}), leading to
\begin{equation}
    \text{CDF}(\rho_{1d}) = \frac{\rho_{1d}v}{N}
\end{equation}
with $\rho_{1d}\in (0,\frac{N}{v}]$. Taking the derivative of the cumulative distribution function, this leads to a uniform probability density function (pdf)
\begin{equation}
\text{pdf}(\rho_{1d})=\frac{v}{N}
\end{equation}

In two dimensions, we consider that the origin is located at the centre of a circular space of radius 1, while the destinations are located uniformly at a distance from the centre $R=1$, and the vessels transit with ballistic motion at speed $v$ from the origin towards a randomly chosen destination. In this case, considering the polar coordinates $r$ and $\phi$, we compute the transit time of a vessel that crosses a region at distance $r$ from the origin of size $\text{d}r \times \text{d} \phi$. Analogously to the one-dimensional system, the transit time of a vessel is $\text{d}r/v$. Additionally, the number of vessels that transit this region is $N \text{d}\phi/(2\pi)$, leading to a total transit time of $\frac{N \text{d}r \text{d}\phi}{2\pi v}$. However, if we consider cells of size $dr\times d\phi$, these cells will be larger for higher $r$, such that the number of vessels on them does not represent the shipping density. To solve this, we consider uniform cells of size $\text{d}x \text{d}y$, with $\text{d}r\text{d}\phi=\frac{\text{d}x\text{d}y}{r}$. Thus, the shipping density, obtained as the transit time per unit area, $\rho_{2d}$:
\begin{equation}
    \rho_{2d} (r,\phi)= \frac{N}{2\pi r v}
    \label{eq:rho2d}
\end{equation}
which, due to the symmetry of the system, does not depend on $\phi$. In this case, the cumulative distribution function will be the fraction of area with density lower than $\rho$. This area will be $\pi R^2-\pi r^2$, and considering from (\ref{eq:rho2d}) $r =  \frac{N}{2\pi \rho_{2d} v}$, we obtain
\begin{equation}
\text{CDF}(\rho_{2d}) = 1-\frac{N^2}{4 \pi^2 \rho_{2d}^2 v^2}
\end{equation}
 with $\rho_{2d} \in [\frac{N}{2\pi v},\infty)$. Taking the derivative of the cumulative distribution function, we obtain the probability density function, described by a power-law
 \begin{equation}
 \text{pdf}(\rho_{2d})=\frac{N^2\rho_{2d}^{-3}}{2\pi^2 v^2}
 \end{equation}

Summing up, a random null model considering an origin-destination flux, with a fixed origin and a uniform probability of reaching any destination or vice-versa, leads to a uniform distribution in one dimension, while the distribution is heavy-tailed with an exponent 3 in two dimensions.

\begin{figure*}[bt!] 
\centering
\includegraphics[width=\textwidth]{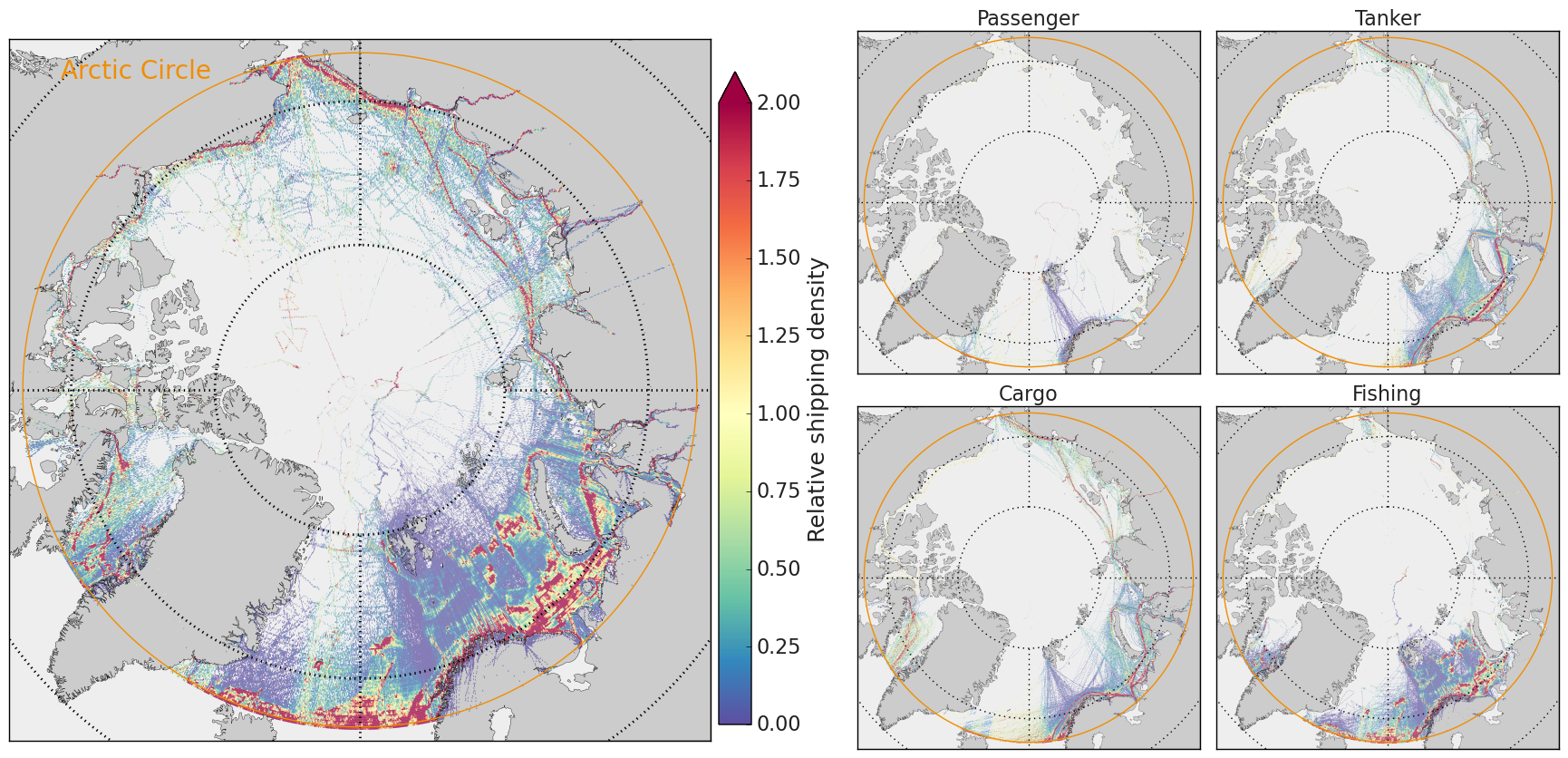}
\caption{Route detection on shipping traffic. For every cell, given its longitude, we computed the relative shipping density as the shipping density (as represented in Fig. \ref{fig:f1}), divided by the average shipping density for that longitude, computed over cells with non-null density. This highlighted some hot spots at each longitude, which we associate with the shipping routes. Left panel represents the total shipping, while right panels are broken down into vessel categories.}\label{fig:f3}
\end{figure*}

To detect shipping routes, we computed the average shipping density per longitude, considering only the grid cells with non-zero values, and represent the relative shipping density of each grid cell, that is, the shipping density divided by the average shipping density of its longitude (see Methods). We observed that the highest values are located in the proximity of the shore (Fig. \ref{fig:f3}). This technique revealed the main shipping routes (red corridors on Fig. \ref{fig:f3}), as well as several fishing hot spots in the high seas. We detected two main Arctic shipping routes, the Northeastern and the Northwest Passage routes, both linking the Northern Pacific and the Northern Atlantic oceans, with the Northeastern Route splitting in two at the North and South of Lyakhovsky Islands. Most of the traffic of both routes was associated with tanker and cargo vessels, as shown by the absence of spatial continuity along the routes in passenger and fishing vessels, which are expected to display shorter range trips.

\begin{figure}[bt!] 
\centering
\includegraphics[width=0.48\textwidth]{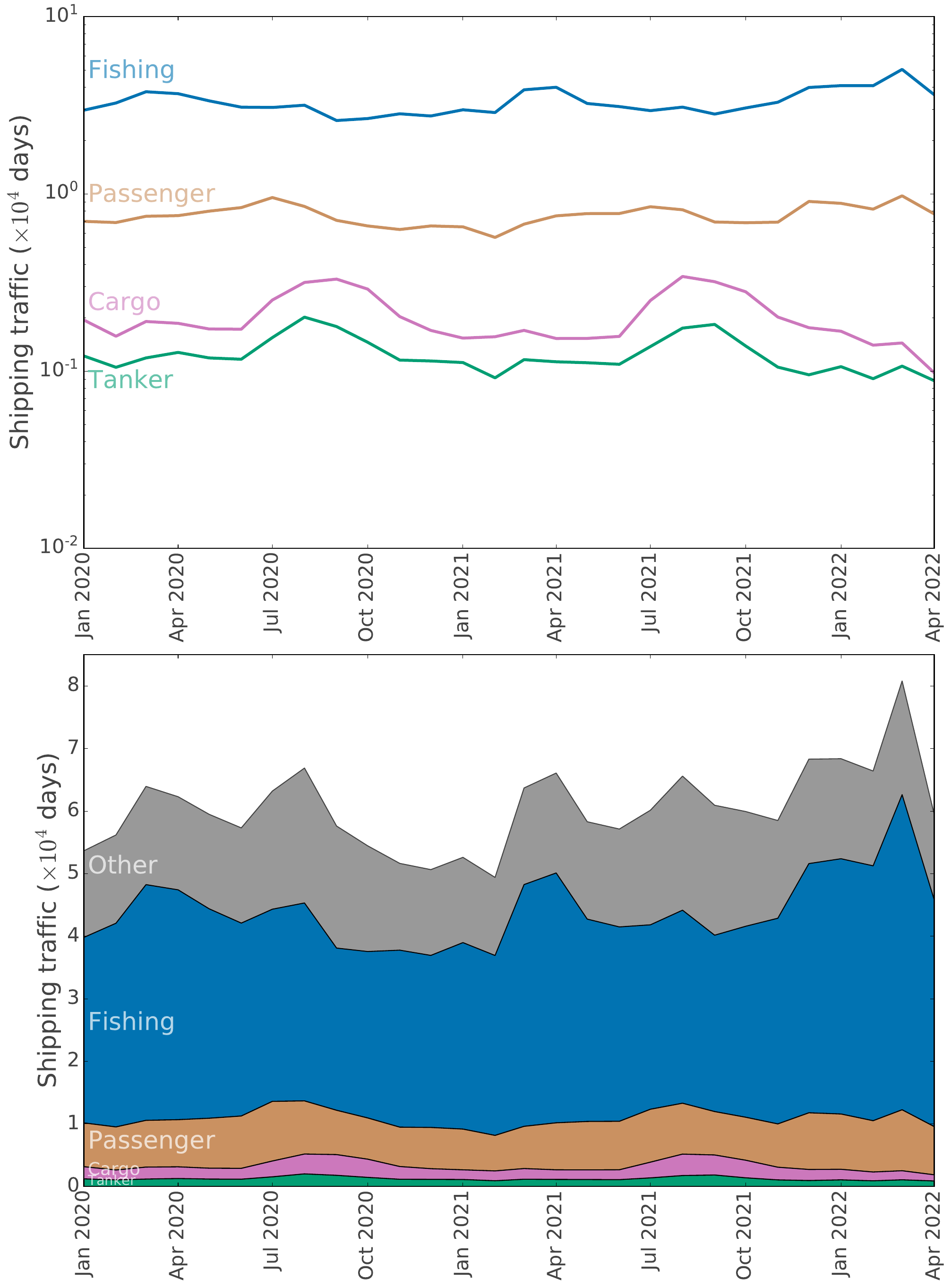}
\caption{Time evolution of the total shipping traffic through the Arctic Ocean. The top panel represents the independent evolution of each vessel category, while the bottom panel shows the joint evolution across different categories.}\label{fig:f4}
\end{figure}

After analyzing the spatial properties of the shipping density at the aggregated dataset and across different vessels categories, we focused on the temporal evolution of the shipping traffic, aggregating all the observations across the Arctic Circle. The highest shipping traffic in the whole period corresponded to fishing vessels followed, in decreasing order, by passenger, cargo and tanker vessels (Fig. \ref{fig:f4}). While the relative evolution of fishing and passenger vessels did not display large relative fluctuations, the shipping traffic for cargo and tanker vessels showed a maximum activity in the summer and early autumn of both 2020 and 2021. 

\begin{figure}[bt!]
\centering
\includegraphics[width=0.49\textwidth]{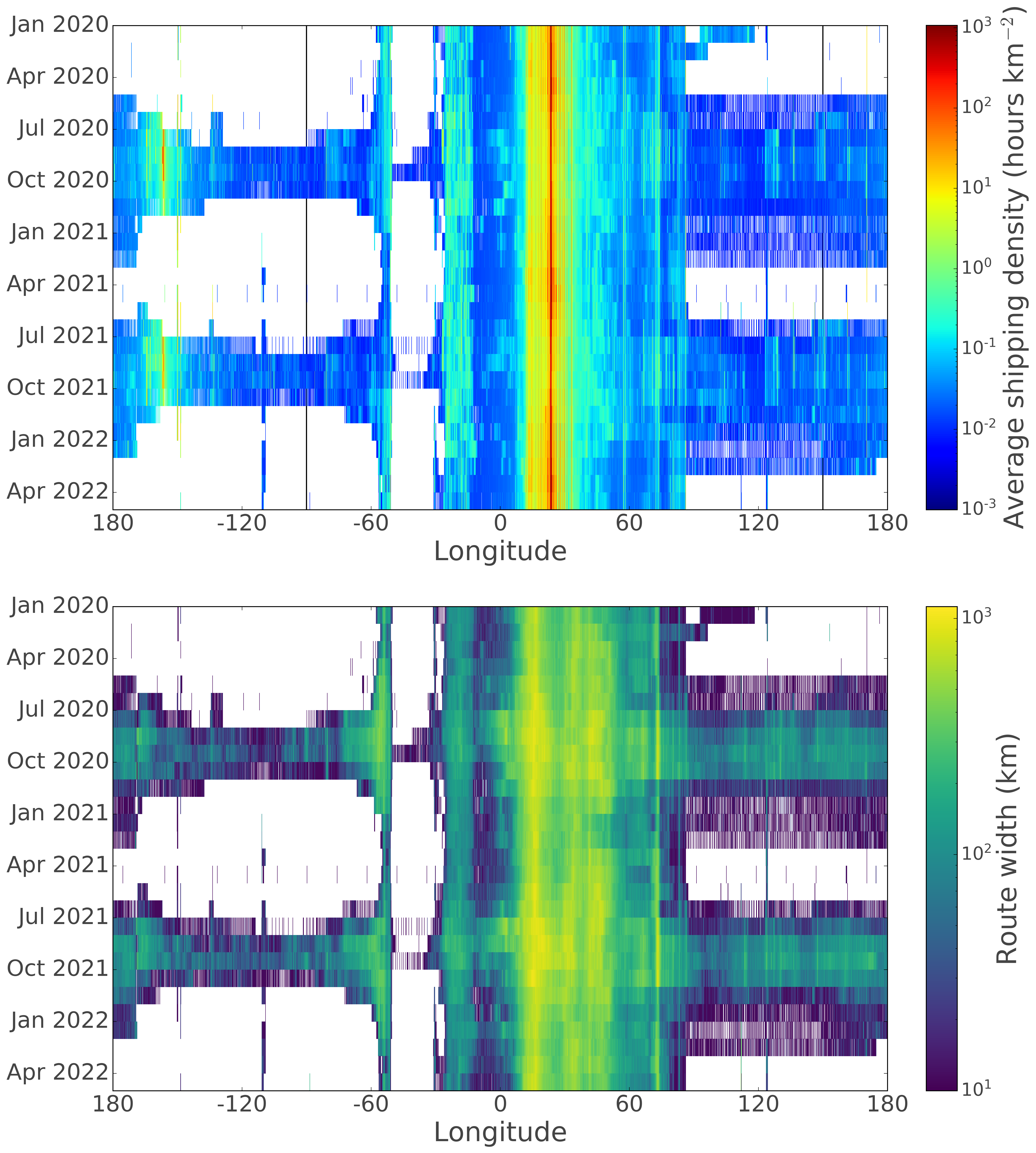}
\caption{Time evolution of the shipping traffic per longitude. The top panel represents the average shipping density over cells with non-null shipping at each longitude, while the bottom represents the sectional length of these cells. White entries represent the absence of traffic. The black lines in the top panel stand for the longitudes that we chose as the most representative for measuring the Northeastern and the Northwest Passage routes being, respectively, 150º and -90º.}\label{fig:f5}
\end{figure}

\begin{figure}[bt!]
\centering
\includegraphics[width=0.49\textwidth]{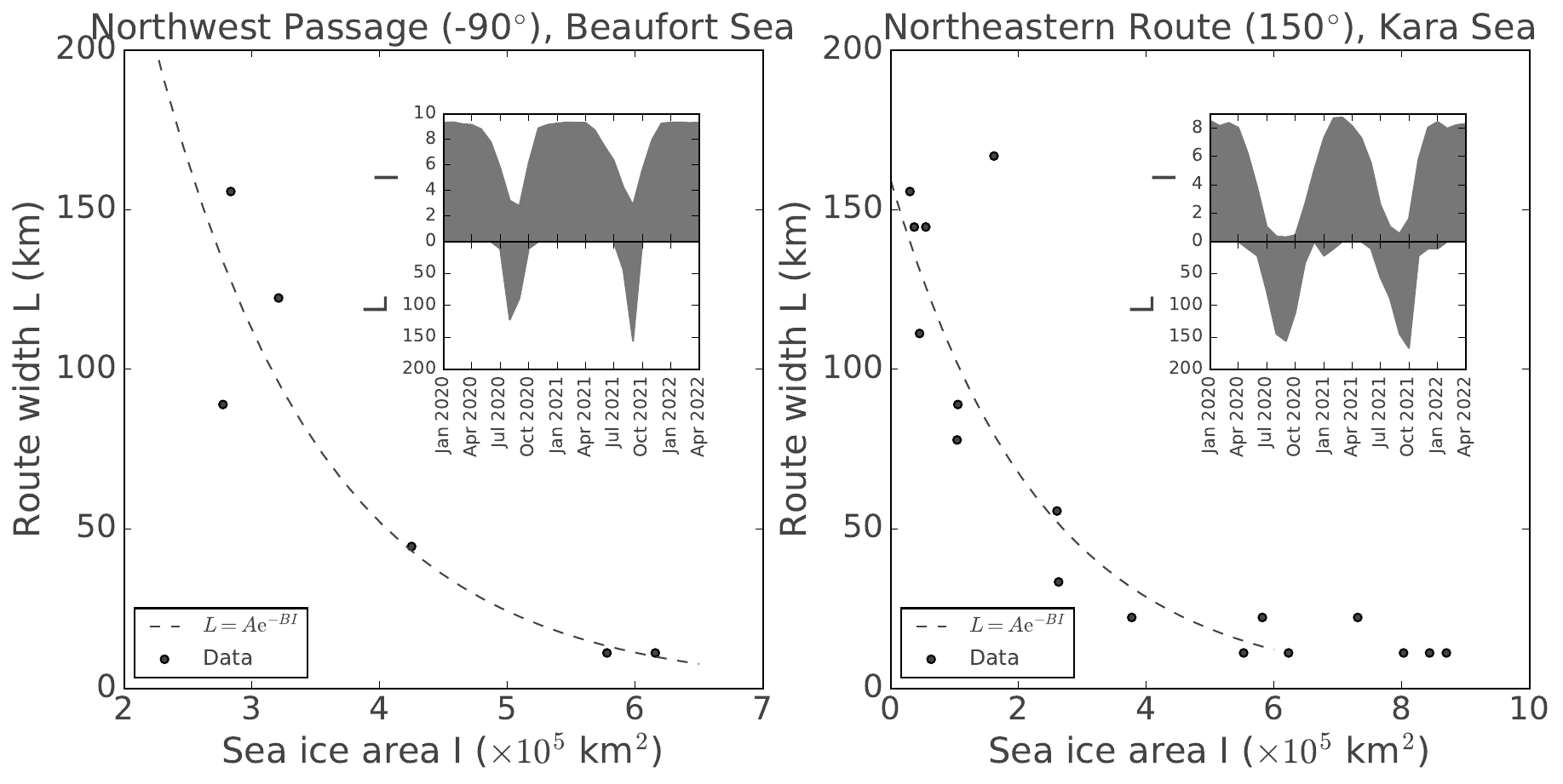}
\caption{Correlation between traffic and sea ice area. We selected two specific longitudes that displayed a seasonal behavior: -90º (West, Northwest Passage route) and 150º (East, Northeastern route), computing the route width $L$ (Fig. \ref{fig:f5}). We computed the Pearson correlation of the route width with the non-zero values of the Sea Ice Area in different Arctic seas, obtaining the maximum absolute value of the correlation for the Beaufort Sea (West, $C=-0.90$) and Kara Sea (East, $C=-0.84$). The main plots represent the route width as a function of the sea ice area, while the insets depict their temporal evolution. Points represent data, while the dashed lines are exponentially decreasing fits, $L=A\mathrm{e}^{-BI}$, with $B_{\mathrm{NWP}}=7.68 \times 10^{-6}$ km$^{-2}$ and $B_{\mathrm{NER}}= 4.29\times 10^{-6}$ km$^{-2}$ for, respectively, the Northwest Passage and the Northeastern Route.}
\label{fig:f6}
\end{figure}

To quantify the time evolution of the shipping routes that we observed in Fig. \ref{fig:f3}, we computed the average shipping density and the route width for each meridional cross-section (that is, zones with constant longitude and varying latitude, see Methods), leading to a representation of the zonal variability of these variables. First, we observed that the average shipping density displayed a seasonal behavior in most of the longitudes except for those where Norwegian, Greenland and Barents seas are, where the traffic was not interrupted in the analyzed period (Fig. \ref{fig:f5}, top). We also quantified the time evolution across longitudes of the route width, observing the same seasonal behaviour, and the appearance of long-range routes whose widths were maximum at the late summer and early autumn, with vessels using the the Northeastern route for four months, while they transited the Northwest Passage route for 2-3 months (Fig. \ref{fig:f5}, bottom), with these periods overlapping with the relative maxima of the shipping traffic for cargo and tanker vessels (Fig. \ref{fig:f4}, top). This temporal evolution of the routes allowed us to select two longitudes as the most representative for the Northeastern (150$^{\circ}$) and the Northwest Passage (-90$^{\circ}$) routes. We considered that the environmental variable with the highest impact on these two shipping routes is the sea ice area, and assessed the correlations between this variable and the route width on the selected longitudes of -90$^{\circ}$ and 150$^{\circ}$. We linked the sea ice area in Baffin Bay, Canadian Archipelago and Beaufort Sea to the Northwest Passage route, and the sea ice area in East Siberian, Kara and Barents seas to the Northeastern route. We checked the correlations between the route width and the sea ice area in these regions, considering only the non-zero route widths, obtaining negative values in all the cases (i.e., as expected, an increase in sea ice area led to a decrease in the vessel traffic route width). In the case of the Northwest Passage route, the absolute correlation between the vessels route width at -90$^{\circ}$ longitude and the sea ice area was maximum for the Beaufort Sea ($C=-0.90$), with correlations of $C=-0.78$ and $C=-0.72$ for, respectively, the Canadian Archipelago and the Baffin Bay. On the other hand, for the Northeastern route, the absolute correlation between the vessels' route width at 150$^{\circ}$ longitude and the ice area displayed its maximum value for the Kara Sea ($C=-0.842$), showing the correlations of $C=-0.837$ and $C=-0.67$ for, respectively, the East Siberian Sea and the Barents Sea. We focused on the areas with maximum absolute correlation between the sea ice area and the route width for the Northwest Passage and the Northeastern routes, assessing the functional relationship between these variables finding, as suggested by the negative correlations, that the route width was maximum when the sea ice area was minimum (Fig. \ref{fig:f6}, inset), and obtaining an exponential decrease of the route width with the sea ice area, with characteristic areas of $1.3\times10^5$ km$^2$ and $2.3\times 10^5$ km$^2$ for the sea ice area in, respectively, the Beaufort Sea and the Kara Sea (Fig. \ref{fig:f6}).

\begin{figure}[bt!]
\centering
\includegraphics[width=0.49\textwidth]{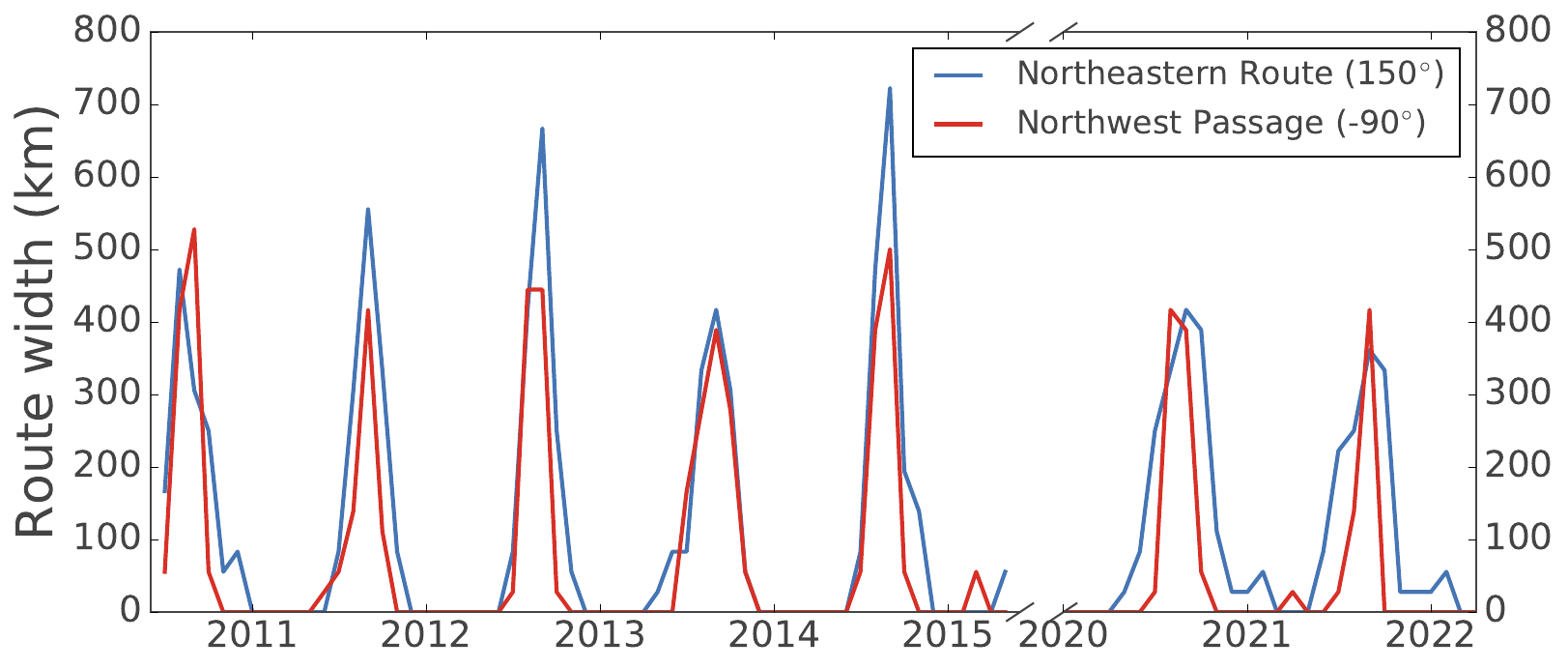}
\caption{Time evolution of the route width at two specific longitudes: -90º (West, Northwest Passage route) and 150º (East, Northeastern route), to characterize the long-term evolution of shipping traffic. The local route width is computed as, for each constant longitude and each month, the cross-sectional length of the cells that display non-null shipping traffic. Data between 2010 and 2015 has been extracted from a different database \citep{eguiluz2016quantitative} with resolution of 0.25$^{\circ}$, number of unique vessels per month. For comparability, this figure includes our analyzed dataset considering grid cells of 0.25$^{\circ}$ side instead of those of 0.1º side.}
\label{fig:f7}
\end{figure}

Finally, we quantified the long-term evolution of the Northwest Passage and the Northeastern routes with a complementary dataset that described the number of unique vessels observed at each 0.25$^{\circ} \times$ 0.25$^{\circ}$ grid cell between July 2010 and May 2015. To warrant the comparability with our dataset, we obtained the vessel route width following the same procedure as described above, but considering the same grid cell size, that is, cells with 0.25$^{\circ}$ side. We observed a similar pattern across different years in the case of the Northwest Passage, while the Northeastern route displayed a remarkable reduction in its maximum value on the most recent dataset, together with a growth in the period when the route displayed observable activity (Fig. \ref{fig:f7}).

\section{Discussion}

Data availability on the marine environment, for example animal trajectories or threats to the marine life \cite{sequeira2019overhauling}, has been historically limited in contrast to the data describing processes occurring on land. However, a change of paradigm from the community and the scientific funding agencies towards data sharing policies is translated into more frequent marine data releases. In this context, online platforms gathering, cleaning, and standardizing multiple source datasets represent a major benefit to advance marine science \cite{tanhua2019ocean,buck2019ocean}. Moreover, the availability of servers for online computation brings a shift towards the democratization of not only the data access, but also the computational resources to process big data. In this context, we have developed our analysis of the Arctic shipping traffic at no cost using the Ocean Data Connector of HUB Ocean, in a Private Preview Week.

We observed that the shipping density was broadly distributed across the Arctic region, with a few locations with a shipping density that was several orders of magnitude higher than the average shipping densities, revealing hot spots of shipping transit (Figs. \ref{fig:f1}, \ref{fig:f2}). We proposed two null models of shipping traffic to understand the observed patterns, obtaining analytically that uniform distributions were associated with one-dimensional traffic, while power-law distributions with exponent $\alpha_{2D}=3$ represented two-dimensional traffic. Although the observed exponents (Fig. \ref{fig:f2}) were different from these two null models, the models allowed us to understand the variability of these exponents across different vessel categories. Particularly, fishing vessels displayed the highest exponent ($\alpha_{\text{fishing}}=1.96$) suggesting, in parallelism to the two-dimensional null model, traffic between different fishing ports and the closest fisheries.  In contrast, passenger vessels showed the lowest exponent $\alpha_{\text{passenger}}=1.49$, which we associate with more directed routes between a few ports, displaying some heterogeneity in the shipping density due to environmental and political factors such as the weather, the ice distribution, Exclusive Economic Zones boundaries or oceanic currents. Finally, cargo and tanker vessels shipping density distributions were described by intermediate exponents ($\alpha_{\text{cargo}}=1.74$, $\alpha_{\text{tanker}=1.90}$), with the latter being higher, which we link to the higher space occupancy of tanker vessels in the Barents Sea, where cargo vessels displayed more concentrated trajectories (Fig. \ref{fig:f1}).

Our analysis of the connection between the shipping traffic route width and the sea ice area revealed maximum absolute (negative) correlations with the sea ice area in regions located far from the locations that we selected as representative for the Northwest Passage and the Northeastern routes. Specifically, for the Northeastern route, we selected the longitude $\lambda_{NE}=150^{\circ}$, which crosses the East Siberian Sea but, although we considered the sea ice area in that sea, the correlation was higher with the sea ice area in the Kara Sea. Similarly, we selected $\lambda_{NWP}=-90^{\circ}$ to analyse the route width of the Northwest Passage route, crossing the area of the Canadian Archipelago but leading to a higher correlation with the sea ice area in the Beaufort Sea. This suggests the presence of ice bottlenecks where windows of ice-free conditions enable the opening of long-distance routes, such that far locations display a correlation with the sea ice area on these ice bottlenecks.

The long-term analysis of the shipping traffic on the Northeastern and the Northwest Passage routes reported an expected seasonal pattern in both routes, displaying a low variability on the Northwest Passage. However, the Northeastern Route showed a recent reduction on the maximum route width, together with a longer seasonal behaviour, \emph{i.e.} the route was narrower, but used through a longer fraction of the year. The local, regional and international mobility restrictions to reduce the spread of COVID-19, on the already called ``anthropause'' \cite{rutz2020covid,bates2021global}, overlapped with most of our analyzed period, implying as well a reduction on factory production rates, and may explain this decrease in the width of the Northeastern route, as a decrease in traffic would lead to vessels following paths that are closer to the optimal, in terms of distance, considering all the geographical and environmental (ice) constraints. In fact, a global analysis revealed a decline of 1.4\% on traffic occupancy in the early months of the pandemic, especially in the Northern Hemisphere \cite{march2021tracking}. 

The decline in the extent of sea ice in the Arctic Ocean with rapid Arctic warming represents an opportunity to optimize shipping routes, reducing the transit duration and costs and, therefore, the greenhouse gas emissions. However, this increase in Arctic shipping traffic may be a harbinger of the ``blue acceleration'' \cite{jouffray2020blue} on the Arctic, threatening the marine species that were not previously exposed to shipping hazards \cite{burek2008effects}, such as ship strikes \cite{reeves2012implications} or underwater noise \cite{tervo2023stuck}, in addition to other stressors such as the already detected plastic pollution \cite{cozar2017arctic,peeken2018arctic}. On the other hand, the availability of new shipping routes may represent a human-based negative feedback on global warming, reducing the emissions from vessels due to the use of shorter routes \cite{browse2013impact}. These potential positive and negative aspects highlight the importance of data analyses to monitor and manage Arctic shipping traffic, conducive to minimize environmental impacts.

\section{Methods}

{\bf Shipping density.} The raw data provided the shipping transit time on high-resolution grid cells of vessels equipped with AIS devices, including all the transiting vessels and its breakdown into four vessel categories, fishing, passenger, cargo and tanker. We aggregated these high-resolution grid cells onto grid cells of size 0.1$^{\circ}\times$ 0.1$^{\circ}$. We obtained the shipping density dividing the shipping transit time by the cell area $A$:

\begin{equation}
    A = R^2 \left(\sin \frac{\pi (\phi+\Delta \phi)}{180}-\sin\frac{\pi \phi}{180}\right) \frac{\Delta \lambda \pi}{180}
    \label{eq:eqarea}
\end{equation}
where $R=6371$ km is the Earth radius, $\phi$ is the latitude in degrees, and $\Delta \lambda$ and $\Delta \phi$ are, respectively, the longitudinal and latitudinal sides of the cell in degrees.

{\bf Average shipping density per longitude.} We compute the average shipping density per longitude as the sum of the total shipping transit times through cells with the same longitude, divided by the area of those cells with a non-zero transit time, following Eq. \ref{eq:eqarea}. For the time evolution of this value (Fig. \ref{fig:f5}, top), at each time step we only consider the cells with non-zero transit time on that specific period.

{\bf Route width per longitude.} We consider all the grid cells in the Arctic region with a specific longitude $\lambda$, and compute the time evolution of the number of cells $N(\lambda,t)$ that displayed a non-zero transit time. We computed the route width $W(\lambda,t)$ as the length of the latitudinal cross-section of these cells:
\begin{equation}
    W(\lambda,t) = R N(\lambda,t) \frac{\Delta \phi \pi}{180}
\end{equation}

\section{Declarations}

\subsection{Funding}

J.P.R. was supported by Juan de la Cierva Formacion program (Ref. FJC2019-040622-I) funded by MCIN/AEI/ 10.13039/501100011033. J.P.R. received support from Spanish Research Agency MCIN/AEI/10.13039/501100011033 via project MISLAND (PID2020-114324GB-C22).

This research is supported by María de Maeztu Excellence Unit 2023-2027 Refs. CEX2021-001201-M and CEX2021-001164-M, funded by MCIN/AEI /10.13039/501100011033.
\subsection{Author's Contributions}

JPR: Conceptualization, Data Curation, Formal Analysis, Software, Visualization, Writing: Original Draft. KK: Formal Analysis, Writing: Review \& Editing. CMD: Conceptualization, Writing: Review \& Editing. VME: Conceptualization, Formal Analysis, Writing: Original Draft.

\section{Acknowledgements}

The authors acknowledge the platform HUB Ocean \url{hubocean.earth} for the access to the data and the computational facilities to remotely run all the analyses, through the Ocean Data Connector.


\begin{thebibliography}{25}%
\makeatletter
\providecommand \@ifxundefined [1]{%
 \@ifx{#1\undefined}
}%
\providecommand \@ifnum [1]{%
 \ifnum #1\expandafter \@firstoftwo
 \else \expandafter \@secondoftwo
 \fi
}%
\providecommand \@ifx [1]{%
 \ifx #1\expandafter \@firstoftwo
 \else \expandafter \@secondoftwo
 \fi
}%
\providecommand \natexlab [1]{#1}%
\providecommand \enquote  [1]{``#1''}%
\providecommand \bibnamefont  [1]{#1}%
\providecommand \bibfnamefont [1]{#1}%
\providecommand \citenamefont [1]{#1}%
\providecommand \href@noop [0]{\@secondoftwo}%
\providecommand \href [0]{\begingroup \@sanitize@url \@href}%
\providecommand \@href[1]{\@@startlink{#1}\@@href}%
\providecommand \@@href[1]{\endgroup#1\@@endlink}%
\providecommand \@sanitize@url [0]{\catcode `\\12\catcode `\$12\catcode `\&12\catcode `\#12\catcode `\^12\catcode `\_12\catcode `\%12\relax}%
\providecommand \@@startlink[1]{}%
\providecommand \@@endlink[0]{}%
\providecommand \url  [0]{\begingroup\@sanitize@url \@url }%
\providecommand \@url [1]{\endgroup\@href {#1}{\urlprefix }}%
\providecommand \urlprefix  [0]{URL }%
\providecommand \Eprint [0]{\href }%
\providecommand \doibase [0]{http://dx.doi.org/}%
\providecommand \selectlanguage [0]{\@gobble}%
\providecommand \bibinfo  [0]{\@secondoftwo}%
\providecommand \bibfield  [0]{\@secondoftwo}%
\providecommand \translation [1]{[#1]}%
\providecommand \BibitemOpen [0]{}%
\providecommand \bibitemStop [0]{}%
\providecommand \bibitemNoStop [0]{.\EOS\space}%
\providecommand \EOS [0]{\spacefactor3000\relax}%
\providecommand \BibitemShut  [1]{\csname bibitem#1\endcsname}%
\let\auto@bib@innerbib\@empty
\bibitem [{\citenamefont {UNCTAD}(2021)}]{unctad2021rev}%
  \BibitemOpen
  \bibfield  {author} {\bibinfo {author} {\bibnamefont {UNCTAD}},\ }\href@noop {} {\emph {\bibinfo {title} {Review of {M}aritime {T}ransport}}}\ (\bibinfo  {publisher} {United Nations Publication},\ \bibinfo {year} {2021})\BibitemShut {NoStop}%
\bibitem [{\citenamefont {Tournadre}(2014)}]{tournadre2014anthropogenic}%
  \BibitemOpen
  \bibfield  {author} {\bibinfo {author} {\bibfnamefont {J.}~\bibnamefont {Tournadre}},\ }\href@noop {} {\bibfield  {journal} {\bibinfo  {journal} {Geophysical Research Letters}\ }\textbf {\bibinfo {volume} {41}},\ \bibinfo {pages} {7924} (\bibinfo {year} {2014})}\BibitemShut {NoStop}%
\bibitem [{\citenamefont {Melia}\ \emph {et~al.}(2016)\citenamefont {Melia}, \citenamefont {Haines},\ and\ \citenamefont {Hawkins}}]{Melia2016}%
  \BibitemOpen
  \bibfield  {author} {\bibinfo {author} {\bibfnamefont {N.}~\bibnamefont {Melia}}, \bibinfo {author} {\bibfnamefont {K.}~\bibnamefont {Haines}}, \ and\ \bibinfo {author} {\bibfnamefont {E.}~\bibnamefont {Hawkins}},\ }\href {\doibase https://doi.org/10.1002/2016GL069315} {\bibfield  {journal} {\bibinfo  {journal} {Geophysical Research Letters}\ }\textbf {\bibinfo {volume} {43}},\ \bibinfo {pages} {9720} (\bibinfo {year} {2016})}\BibitemShut {NoStop}%
\bibitem [{\citenamefont {M{\"u}ller}\ \emph {et~al.}(2023)\citenamefont {M{\"u}ller}, \citenamefont {Knol-Kauffman}, \citenamefont {Jeuring},\ and\ \citenamefont {Palerme}}]{muller2023arctic}%
  \BibitemOpen
  \bibfield  {author} {\bibinfo {author} {\bibfnamefont {M.}~\bibnamefont {M{\"u}ller}}, \bibinfo {author} {\bibfnamefont {M.}~\bibnamefont {Knol-Kauffman}}, \bibinfo {author} {\bibfnamefont {J.}~\bibnamefont {Jeuring}}, \ and\ \bibinfo {author} {\bibfnamefont {C.}~\bibnamefont {Palerme}},\ }\href@noop {} {\bibfield  {journal} {\bibinfo  {journal} {npj Ocean Sustainability}\ }\textbf {\bibinfo {volume} {2}},\ \bibinfo {pages} {12} (\bibinfo {year} {2023})}\BibitemShut {NoStop}%
\bibitem [{\citenamefont {Kroodsma}\ \emph {et~al.}(2018)\citenamefont {Kroodsma}, \citenamefont {Mayorga}, \citenamefont {Hochberg}, \citenamefont {Miller}, \citenamefont {Boerder}, \citenamefont {Ferretti}, \citenamefont {Wilson}, \citenamefont {Bergman}, \citenamefont {White}, \citenamefont {Block} \emph {et~al.}}]{kroodsma2018tracking}%
  \BibitemOpen
  \bibfield  {author} {\bibinfo {author} {\bibfnamefont {D.~A.}\ \bibnamefont {Kroodsma}}, \bibinfo {author} {\bibfnamefont {J.}~\bibnamefont {Mayorga}}, \bibinfo {author} {\bibfnamefont {T.}~\bibnamefont {Hochberg}}, \bibinfo {author} {\bibfnamefont {N.~A.}\ \bibnamefont {Miller}}, \bibinfo {author} {\bibfnamefont {K.}~\bibnamefont {Boerder}}, \bibinfo {author} {\bibfnamefont {F.}~\bibnamefont {Ferretti}}, \bibinfo {author} {\bibfnamefont {A.}~\bibnamefont {Wilson}}, \bibinfo {author} {\bibfnamefont {B.}~\bibnamefont {Bergman}}, \bibinfo {author} {\bibfnamefont {T.~D.}\ \bibnamefont {White}}, \bibinfo {author} {\bibfnamefont {B.~A.}\ \bibnamefont {Block}},  \emph {et~al.},\ }\href@noop {} {\bibfield  {journal} {\bibinfo  {journal} {Science}\ }\textbf {\bibinfo {volume} {359}},\ \bibinfo {pages} {904} (\bibinfo {year} {2018})}\BibitemShut {NoStop}%
\bibitem [{\citenamefont {Rodr{\'\i}guez}\ \emph {et~al.}(2021)\citenamefont {Rodr{\'\i}guez}, \citenamefont {Fern{\'a}ndez-Gracia}, \citenamefont {Duarte}, \citenamefont {Irigoien},\ and\ \citenamefont {Egu{\'\i}luz}}]{rodriguez2021global}%
  \BibitemOpen
  \bibfield  {author} {\bibinfo {author} {\bibfnamefont {J.~P.}\ \bibnamefont {Rodr{\'\i}guez}}, \bibinfo {author} {\bibfnamefont {J.}~\bibnamefont {Fern{\'a}ndez-Gracia}}, \bibinfo {author} {\bibfnamefont {C.~M.}\ \bibnamefont {Duarte}}, \bibinfo {author} {\bibfnamefont {X.}~\bibnamefont {Irigoien}}, \ and\ \bibinfo {author} {\bibfnamefont {V.~M.}\ \bibnamefont {Egu{\'\i}luz}},\ }\href@noop {} {\bibfield  {journal} {\bibinfo  {journal} {Science Advances}\ }\textbf {\bibinfo {volume} {7}},\ \bibinfo {pages} {eabe3470} (\bibinfo {year} {2021})}\BibitemShut {NoStop}%
\bibitem [{\citenamefont {Frawley}\ \emph {et~al.}(2022)\citenamefont {Frawley}, \citenamefont {Muhling}, \citenamefont {Welch}, \citenamefont {Seto}, \citenamefont {Chang}, \citenamefont {Blaha}, \citenamefont {Hanich}, \citenamefont {Jung}, \citenamefont {Hazen}, \citenamefont {Jacox} \emph {et~al.}}]{frawley2022clustering}%
  \BibitemOpen
  \bibfield  {author} {\bibinfo {author} {\bibfnamefont {T.~H.}\ \bibnamefont {Frawley}}, \bibinfo {author} {\bibfnamefont {B.}~\bibnamefont {Muhling}}, \bibinfo {author} {\bibfnamefont {H.}~\bibnamefont {Welch}}, \bibinfo {author} {\bibfnamefont {K.~L.}\ \bibnamefont {Seto}}, \bibinfo {author} {\bibfnamefont {S.-K.}\ \bibnamefont {Chang}}, \bibinfo {author} {\bibfnamefont {F.}~\bibnamefont {Blaha}}, \bibinfo {author} {\bibfnamefont {Q.}~\bibnamefont {Hanich}}, \bibinfo {author} {\bibfnamefont {M.}~\bibnamefont {Jung}}, \bibinfo {author} {\bibfnamefont {E.~L.}\ \bibnamefont {Hazen}}, \bibinfo {author} {\bibfnamefont {M.~G.}\ \bibnamefont {Jacox}},  \emph {et~al.},\ }\href@noop {} {\bibfield  {journal} {\bibinfo  {journal} {One Earth}\ }\textbf {\bibinfo {volume} {5}},\ \bibinfo {pages} {1002} (\bibinfo {year} {2022})}\BibitemShut {NoStop}%
\bibitem [{\citenamefont {Queiroz}\ \emph {et~al.}(2019)\citenamefont {Queiroz}, \citenamefont {Humphries}, \citenamefont {Couto}, \citenamefont {Vedor}, \citenamefont {Da~Costa}, \citenamefont {Sequeira}, \citenamefont {Mucientes}, \citenamefont {Santos}, \citenamefont {Abascal}, \citenamefont {Abercrombie} \emph {et~al.}}]{queiroz2019global}%
  \BibitemOpen
  \bibfield  {author} {\bibinfo {author} {\bibfnamefont {N.}~\bibnamefont {Queiroz}}, \bibinfo {author} {\bibfnamefont {N.~E.}\ \bibnamefont {Humphries}}, \bibinfo {author} {\bibfnamefont {A.}~\bibnamefont {Couto}}, \bibinfo {author} {\bibfnamefont {M.}~\bibnamefont {Vedor}}, \bibinfo {author} {\bibfnamefont {I.}~\bibnamefont {Da~Costa}}, \bibinfo {author} {\bibfnamefont {A.~M.}\ \bibnamefont {Sequeira}}, \bibinfo {author} {\bibfnamefont {G.}~\bibnamefont {Mucientes}}, \bibinfo {author} {\bibfnamefont {A.~M.}\ \bibnamefont {Santos}}, \bibinfo {author} {\bibfnamefont {F.~J.}\ \bibnamefont {Abascal}}, \bibinfo {author} {\bibfnamefont {D.~L.}\ \bibnamefont {Abercrombie}},  \emph {et~al.},\ }\href@noop {} {\bibfield  {journal} {\bibinfo  {journal} {Nature}\ }\textbf {\bibinfo {volume} {572}},\ \bibinfo {pages} {461} (\bibinfo {year} {2019})}\BibitemShut {NoStop}%
\bibitem [{\citenamefont {Womersley}\ \emph {et~al.}(2022)\citenamefont {Womersley}, \citenamefont {Humphries}, \citenamefont {Queiroz}, \citenamefont {Vedor}, \citenamefont {da~Costa}, \citenamefont {Furtado}, \citenamefont {Tyminski}, \citenamefont {Abrantes}, \citenamefont {Araujo}, \citenamefont {Bach} \emph {et~al.}}]{womersley2022global}%
  \BibitemOpen
  \bibfield  {author} {\bibinfo {author} {\bibfnamefont {F.~C.}\ \bibnamefont {Womersley}}, \bibinfo {author} {\bibfnamefont {N.~E.}\ \bibnamefont {Humphries}}, \bibinfo {author} {\bibfnamefont {N.}~\bibnamefont {Queiroz}}, \bibinfo {author} {\bibfnamefont {M.}~\bibnamefont {Vedor}}, \bibinfo {author} {\bibfnamefont {I.}~\bibnamefont {da~Costa}}, \bibinfo {author} {\bibfnamefont {M.}~\bibnamefont {Furtado}}, \bibinfo {author} {\bibfnamefont {J.~P.}\ \bibnamefont {Tyminski}}, \bibinfo {author} {\bibfnamefont {K.}~\bibnamefont {Abrantes}}, \bibinfo {author} {\bibfnamefont {G.}~\bibnamefont {Araujo}}, \bibinfo {author} {\bibfnamefont {S.~S.}\ \bibnamefont {Bach}},  \emph {et~al.},\ }\href@noop {} {\bibfield  {journal} {\bibinfo  {journal} {Proceedings of the National Academy of Sciences}\ }\textbf {\bibinfo {volume} {119}},\ \bibinfo {pages} {e2117440119} (\bibinfo {year} {2022})}\BibitemShut {NoStop}%
\bibitem [{\citenamefont {Egu{\'\i}luz}\ \emph {et~al.}(2016)\citenamefont {Egu{\'\i}luz}, \citenamefont {Fern{\'a}ndez-Gracia}, \citenamefont {Irigoien},\ and\ \citenamefont {Duarte}}]{eguiluz2016quantitative}%
  \BibitemOpen
  \bibfield  {author} {\bibinfo {author} {\bibfnamefont {V.~M.}\ \bibnamefont {Egu{\'\i}luz}}, \bibinfo {author} {\bibfnamefont {J.}~\bibnamefont {Fern{\'a}ndez-Gracia}}, \bibinfo {author} {\bibfnamefont {X.}~\bibnamefont {Irigoien}}, \ and\ \bibinfo {author} {\bibfnamefont {C.~M.}\ \bibnamefont {Duarte}},\ }\href@noop {} {\bibfield  {journal} {\bibinfo  {journal} {Scientific Reports}\ }\textbf {\bibinfo {volume} {6}},\ \bibinfo {pages} {30682} (\bibinfo {year} {2016})}\BibitemShut {NoStop}%
\bibitem [{\citenamefont {Fetterer}\ \emph {et~al.}(2017)\citenamefont {Fetterer}, \citenamefont {Knowles}, \citenamefont {Meier}, \citenamefont {M.~Savoie},\ and\ \citenamefont {Windnagel}}]{seaice}%
  \BibitemOpen
  \bibfield  {author} {\bibinfo {author} {\bibfnamefont {F.}~\bibnamefont {Fetterer}}, \bibinfo {author} {\bibfnamefont {K.}~\bibnamefont {Knowles}}, \bibinfo {author} {\bibfnamefont {W.~N.}\ \bibnamefont {Meier}}, \bibinfo {author} {\bibfnamefont {M.}~\bibnamefont {M.~Savoie}}, \ and\ \bibinfo {author} {\bibfnamefont {A.~K.}\ \bibnamefont {Windnagel}},\ }\href {\doibase 10.7265/N5K072F8} {\enquote {\bibinfo {title} {Sea {I}ce {I}ndex, version 3},}\ } (\bibinfo {year} {2017})\BibitemShut {NoStop}%
\bibitem [{\citenamefont {Alstott}\ \emph {et~al.}(2014)\citenamefont {Alstott}, \citenamefont {Bullmore},\ and\ \citenamefont {Plenz}}]{alstott2014powerlaw}%
  \BibitemOpen
  \bibfield  {author} {\bibinfo {author} {\bibfnamefont {J.}~\bibnamefont {Alstott}}, \bibinfo {author} {\bibfnamefont {E.}~\bibnamefont {Bullmore}}, \ and\ \bibinfo {author} {\bibfnamefont {D.}~\bibnamefont {Plenz}},\ }\href@noop {} {\bibfield  {journal} {\bibinfo  {journal} {PLoS ONE}\ }\textbf {\bibinfo {volume} {9}},\ \bibinfo {pages} {e85777} (\bibinfo {year} {2014})}\BibitemShut {NoStop}%
\bibitem [{\citenamefont {Sequeira}\ \emph {et~al.}(2019)\citenamefont {Sequeira}, \citenamefont {Hays}, \citenamefont {Sims}, \citenamefont {Egu{\'\i}luz}, \citenamefont {Rodr{\'\i}guez}, \citenamefont {Heupel}, \citenamefont {Harcourt}, \citenamefont {Calich}, \citenamefont {Queiroz}, \citenamefont {Costa} \emph {et~al.}}]{sequeira2019overhauling}%
  \BibitemOpen
  \bibfield  {author} {\bibinfo {author} {\bibfnamefont {A.~M.~M.}\ \bibnamefont {Sequeira}}, \bibinfo {author} {\bibfnamefont {G.~C.}\ \bibnamefont {Hays}}, \bibinfo {author} {\bibfnamefont {D.~W.}\ \bibnamefont {Sims}}, \bibinfo {author} {\bibfnamefont {V.~M.}\ \bibnamefont {Egu{\'\i}luz}}, \bibinfo {author} {\bibfnamefont {J.~P.}\ \bibnamefont {Rodr{\'\i}guez}}, \bibinfo {author} {\bibfnamefont {M.~R.}\ \bibnamefont {Heupel}}, \bibinfo {author} {\bibfnamefont {R.}~\bibnamefont {Harcourt}}, \bibinfo {author} {\bibfnamefont {H.}~\bibnamefont {Calich}}, \bibinfo {author} {\bibfnamefont {N.}~\bibnamefont {Queiroz}}, \bibinfo {author} {\bibfnamefont {D.~P.}\ \bibnamefont {Costa}},  \emph {et~al.},\ }\href@noop {} {\bibfield  {journal} {\bibinfo  {journal} {Frontiers in Marine Science}\ }\textbf {\bibinfo {volume} {6}},\ \bibinfo {pages} {639} (\bibinfo {year} {2019})}\BibitemShut {NoStop}%
\bibitem [{\citenamefont {Tanhua}\ \emph {et~al.}(2019)\citenamefont {Tanhua}, \citenamefont {Pouliquen}, \citenamefont {Hausman}, \citenamefont {O’brien}, \citenamefont {Bricher}, \citenamefont {De~Bruin}, \citenamefont {Buck}, \citenamefont {Burger}, \citenamefont {Carval}, \citenamefont {Casey} \emph {et~al.}}]{tanhua2019ocean}%
  \BibitemOpen
  \bibfield  {author} {\bibinfo {author} {\bibfnamefont {T.}~\bibnamefont {Tanhua}}, \bibinfo {author} {\bibfnamefont {S.}~\bibnamefont {Pouliquen}}, \bibinfo {author} {\bibfnamefont {J.}~\bibnamefont {Hausman}}, \bibinfo {author} {\bibfnamefont {K.}~\bibnamefont {O’brien}}, \bibinfo {author} {\bibfnamefont {P.}~\bibnamefont {Bricher}}, \bibinfo {author} {\bibfnamefont {T.}~\bibnamefont {De~Bruin}}, \bibinfo {author} {\bibfnamefont {J.~J.}\ \bibnamefont {Buck}}, \bibinfo {author} {\bibfnamefont {E.~F.}\ \bibnamefont {Burger}}, \bibinfo {author} {\bibfnamefont {T.}~\bibnamefont {Carval}}, \bibinfo {author} {\bibfnamefont {K.~S.}\ \bibnamefont {Casey}},  \emph {et~al.},\ }\href@noop {} {\bibfield  {journal} {\bibinfo  {journal} {Frontiers in Marine Science}\ }\textbf {\bibinfo {volume} {6}},\ \bibinfo {pages} {440} (\bibinfo {year} {2019})}\BibitemShut {NoStop}%
\bibitem [{\citenamefont {Buck}\ \emph {et~al.}(2019)\citenamefont {Buck}, \citenamefont {Bainbridge}, \citenamefont {Burger}, \citenamefont {Kraberg}, \citenamefont {Casari}, \citenamefont {Casey}, \citenamefont {Darroch}, \citenamefont {Rio}, \citenamefont {Metfies}, \citenamefont {Delory} \emph {et~al.}}]{buck2019ocean}%
  \BibitemOpen
  \bibfield  {author} {\bibinfo {author} {\bibfnamefont {J.~J.}\ \bibnamefont {Buck}}, \bibinfo {author} {\bibfnamefont {S.~J.}\ \bibnamefont {Bainbridge}}, \bibinfo {author} {\bibfnamefont {E.~F.}\ \bibnamefont {Burger}}, \bibinfo {author} {\bibfnamefont {A.~C.}\ \bibnamefont {Kraberg}}, \bibinfo {author} {\bibfnamefont {M.}~\bibnamefont {Casari}}, \bibinfo {author} {\bibfnamefont {K.~S.}\ \bibnamefont {Casey}}, \bibinfo {author} {\bibfnamefont {L.}~\bibnamefont {Darroch}}, \bibinfo {author} {\bibfnamefont {J.~D.}\ \bibnamefont {Rio}}, \bibinfo {author} {\bibfnamefont {K.}~\bibnamefont {Metfies}}, \bibinfo {author} {\bibfnamefont {E.}~\bibnamefont {Delory}},  \emph {et~al.},\ }\href@noop {} {\bibfield  {journal} {\bibinfo  {journal} {Frontiers in Marine Science}\ }\textbf {\bibinfo {volume} {6}},\ \bibinfo {pages} {32} (\bibinfo {year} {2019})}\BibitemShut {NoStop}%
\bibitem [{\citenamefont {Rutz}\ \emph {et~al.}(2020)\citenamefont {Rutz}, \citenamefont {Loretto}, \citenamefont {Bates}, \citenamefont {Davidson}, \citenamefont {Duarte}, \citenamefont {Jetz}, \citenamefont {Johnson}, \citenamefont {Kato}, \citenamefont {Kays}, \citenamefont {Mueller} \emph {et~al.}}]{rutz2020covid}%
  \BibitemOpen
  \bibfield  {author} {\bibinfo {author} {\bibfnamefont {C.}~\bibnamefont {Rutz}}, \bibinfo {author} {\bibfnamefont {M.-C.}\ \bibnamefont {Loretto}}, \bibinfo {author} {\bibfnamefont {A.~E.}\ \bibnamefont {Bates}}, \bibinfo {author} {\bibfnamefont {S.~C.}\ \bibnamefont {Davidson}}, \bibinfo {author} {\bibfnamefont {C.~M.}\ \bibnamefont {Duarte}}, \bibinfo {author} {\bibfnamefont {W.}~\bibnamefont {Jetz}}, \bibinfo {author} {\bibfnamefont {M.}~\bibnamefont {Johnson}}, \bibinfo {author} {\bibfnamefont {A.}~\bibnamefont {Kato}}, \bibinfo {author} {\bibfnamefont {R.}~\bibnamefont {Kays}}, \bibinfo {author} {\bibfnamefont {T.}~\bibnamefont {Mueller}},  \emph {et~al.},\ }\href@noop {} {\bibfield  {journal} {\bibinfo  {journal} {Nature Ecology \& Evolution}\ }\textbf {\bibinfo {volume} {4}},\ \bibinfo {pages} {1156} (\bibinfo {year} {2020})}\BibitemShut {NoStop}%
\bibitem [{\citenamefont {Bates}\ \emph {et~al.}(2021)\citenamefont {Bates}, \citenamefont {Primack}, \citenamefont {Biggar}, \citenamefont {Bird}, \citenamefont {Clinton}, \citenamefont {Command}, \citenamefont {Richards}, \citenamefont {Shellard}, \citenamefont {Geraldi}, \citenamefont {Vergara} \emph {et~al.}}]{bates2021global}%
  \BibitemOpen
  \bibfield  {author} {\bibinfo {author} {\bibfnamefont {A.~E.}\ \bibnamefont {Bates}}, \bibinfo {author} {\bibfnamefont {R.~B.}\ \bibnamefont {Primack}}, \bibinfo {author} {\bibfnamefont {B.~S.}\ \bibnamefont {Biggar}}, \bibinfo {author} {\bibfnamefont {T.~J.}\ \bibnamefont {Bird}}, \bibinfo {author} {\bibfnamefont {M.~E.}\ \bibnamefont {Clinton}}, \bibinfo {author} {\bibfnamefont {R.~J.}\ \bibnamefont {Command}}, \bibinfo {author} {\bibfnamefont {C.}~\bibnamefont {Richards}}, \bibinfo {author} {\bibfnamefont {M.}~\bibnamefont {Shellard}}, \bibinfo {author} {\bibfnamefont {N.~R.}\ \bibnamefont {Geraldi}}, \bibinfo {author} {\bibfnamefont {V.}~\bibnamefont {Vergara}},  \emph {et~al.},\ }\href@noop {} {\bibfield  {journal} {\bibinfo  {journal} {Biological Conservation}\ }\textbf {\bibinfo {volume} {263}},\ \bibinfo {pages} {109175} (\bibinfo {year} {2021})}\BibitemShut {NoStop}%
\bibitem [{\citenamefont {March}\ \emph {et~al.}(2021)\citenamefont {March}, \citenamefont {Metcalfe}, \citenamefont {Tintor{\'e}},\ and\ \citenamefont {Godley}}]{march2021tracking}%
  \BibitemOpen
  \bibfield  {author} {\bibinfo {author} {\bibfnamefont {D.}~\bibnamefont {March}}, \bibinfo {author} {\bibfnamefont {K.}~\bibnamefont {Metcalfe}}, \bibinfo {author} {\bibfnamefont {J.}~\bibnamefont {Tintor{\'e}}}, \ and\ \bibinfo {author} {\bibfnamefont {B.~J.}\ \bibnamefont {Godley}},\ }\href@noop {} {\bibfield  {journal} {\bibinfo  {journal} {Nature Communications}\ }\textbf {\bibinfo {volume} {12}},\ \bibinfo {pages} {2415} (\bibinfo {year} {2021})}\BibitemShut {NoStop}%
\bibitem [{\citenamefont {Jouffray}\ \emph {et~al.}(2020)\citenamefont {Jouffray}, \citenamefont {Blasiak}, \citenamefont {Norstr{\"o}m}, \citenamefont {{\"O}sterblom},\ and\ \citenamefont {Nystr{\"o}m}}]{jouffray2020blue}%
  \BibitemOpen
  \bibfield  {author} {\bibinfo {author} {\bibfnamefont {J.-B.}\ \bibnamefont {Jouffray}}, \bibinfo {author} {\bibfnamefont {R.}~\bibnamefont {Blasiak}}, \bibinfo {author} {\bibfnamefont {A.~V.}\ \bibnamefont {Norstr{\"o}m}}, \bibinfo {author} {\bibfnamefont {H.}~\bibnamefont {{\"O}sterblom}}, \ and\ \bibinfo {author} {\bibfnamefont {M.}~\bibnamefont {Nystr{\"o}m}},\ }\href@noop {} {\bibfield  {journal} {\bibinfo  {journal} {One Earth}\ }\textbf {\bibinfo {volume} {2}},\ \bibinfo {pages} {43} (\bibinfo {year} {2020})}\BibitemShut {NoStop}%
\bibitem [{\citenamefont {Burek}\ \emph {et~al.}(2008)\citenamefont {Burek}, \citenamefont {Gulland},\ and\ \citenamefont {O'Hara}}]{burek2008effects}%
  \BibitemOpen
  \bibfield  {author} {\bibinfo {author} {\bibfnamefont {K.~A.}\ \bibnamefont {Burek}}, \bibinfo {author} {\bibfnamefont {F.~M.}\ \bibnamefont {Gulland}}, \ and\ \bibinfo {author} {\bibfnamefont {T.~M.}\ \bibnamefont {O'Hara}},\ }\href@noop {} {\bibfield  {journal} {\bibinfo  {journal} {Ecological Applications}\ }\textbf {\bibinfo {volume} {18}},\ \bibinfo {pages} {S126} (\bibinfo {year} {2008})}\BibitemShut {NoStop}%
\bibitem [{\citenamefont {Reeves}\ \emph {et~al.}(2012)\citenamefont {Reeves}, \citenamefont {Rosa}, \citenamefont {George}, \citenamefont {Sheffield},\ and\ \citenamefont {Moore}}]{reeves2012implications}%
  \BibitemOpen
  \bibfield  {author} {\bibinfo {author} {\bibfnamefont {R.}~\bibnamefont {Reeves}}, \bibinfo {author} {\bibfnamefont {C.}~\bibnamefont {Rosa}}, \bibinfo {author} {\bibfnamefont {J.~C.}\ \bibnamefont {George}}, \bibinfo {author} {\bibfnamefont {G.}~\bibnamefont {Sheffield}}, \ and\ \bibinfo {author} {\bibfnamefont {M.}~\bibnamefont {Moore}},\ }\href@noop {} {\bibfield  {journal} {\bibinfo  {journal} {Marine policy}\ }\textbf {\bibinfo {volume} {36}},\ \bibinfo {pages} {454} (\bibinfo {year} {2012})}\BibitemShut {NoStop}%
\bibitem [{\citenamefont {Tervo}\ \emph {et~al.}(2023)\citenamefont {Tervo}, \citenamefont {Blackwell}, \citenamefont {Ditlevsen}, \citenamefont {Garde}, \citenamefont {Hansen}, \citenamefont {Samson}, \citenamefont {Conrad},\ and\ \citenamefont {Heide-J{\o}rgensen}}]{tervo2023stuck}%
  \BibitemOpen
  \bibfield  {author} {\bibinfo {author} {\bibfnamefont {O.~M.}\ \bibnamefont {Tervo}}, \bibinfo {author} {\bibfnamefont {S.~B.}\ \bibnamefont {Blackwell}}, \bibinfo {author} {\bibfnamefont {S.}~\bibnamefont {Ditlevsen}}, \bibinfo {author} {\bibfnamefont {E.}~\bibnamefont {Garde}}, \bibinfo {author} {\bibfnamefont {R.~G.}\ \bibnamefont {Hansen}}, \bibinfo {author} {\bibfnamefont {A.~L.}\ \bibnamefont {Samson}}, \bibinfo {author} {\bibfnamefont {A.~S.}\ \bibnamefont {Conrad}}, \ and\ \bibinfo {author} {\bibfnamefont {M.~P.}\ \bibnamefont {Heide-J{\o}rgensen}},\ }\href@noop {} {\bibfield  {journal} {\bibinfo  {journal} {Science Advances}\ }\textbf {\bibinfo {volume} {9}},\ \bibinfo {pages} {eade0440} (\bibinfo {year} {2023})}\BibitemShut {NoStop}%
\bibitem [{\citenamefont {C{\'o}zar}\ \emph {et~al.}(2017)\citenamefont {C{\'o}zar}, \citenamefont {Mart{\'\i}}, \citenamefont {Duarte}, \citenamefont {Garc{\'\i}a-de Lomas}, \citenamefont {Van~Sebille}, \citenamefont {Ballatore}, \citenamefont {Egu{\'\i}luz}, \citenamefont {Gonz{\'a}lez-Gordillo}, \citenamefont {Pedrotti}, \citenamefont {Echevarr{\'\i}a} \emph {et~al.}}]{cozar2017arctic}%
  \BibitemOpen
  \bibfield  {author} {\bibinfo {author} {\bibfnamefont {A.}~\bibnamefont {C{\'o}zar}}, \bibinfo {author} {\bibfnamefont {E.}~\bibnamefont {Mart{\'\i}}}, \bibinfo {author} {\bibfnamefont {C.~M.}\ \bibnamefont {Duarte}}, \bibinfo {author} {\bibfnamefont {J.}~\bibnamefont {Garc{\'\i}a-de Lomas}}, \bibinfo {author} {\bibfnamefont {E.}~\bibnamefont {Van~Sebille}}, \bibinfo {author} {\bibfnamefont {T.~J.}\ \bibnamefont {Ballatore}}, \bibinfo {author} {\bibfnamefont {V.~M.}\ \bibnamefont {Egu{\'\i}luz}}, \bibinfo {author} {\bibfnamefont {J.~I.}\ \bibnamefont {Gonz{\'a}lez-Gordillo}}, \bibinfo {author} {\bibfnamefont {M.~L.}\ \bibnamefont {Pedrotti}}, \bibinfo {author} {\bibfnamefont {F.}~\bibnamefont {Echevarr{\'\i}a}},  \emph {et~al.},\ }\href@noop {} {\bibfield  {journal} {\bibinfo  {journal} {Science Advances}\ }\textbf {\bibinfo {volume} {3}},\ \bibinfo {pages} {e1600582} (\bibinfo {year} {2017})}\BibitemShut {NoStop}%
\bibitem [{\citenamefont {Peeken}\ \emph {et~al.}(2018)\citenamefont {Peeken}, \citenamefont {Primpke}, \citenamefont {Beyer}, \citenamefont {G{\"u}termann}, \citenamefont {Katlein}, \citenamefont {Krumpen}, \citenamefont {Bergmann}, \citenamefont {Hehemann},\ and\ \citenamefont {Gerdts}}]{peeken2018arctic}%
  \BibitemOpen
  \bibfield  {author} {\bibinfo {author} {\bibfnamefont {I.}~\bibnamefont {Peeken}}, \bibinfo {author} {\bibfnamefont {S.}~\bibnamefont {Primpke}}, \bibinfo {author} {\bibfnamefont {B.}~\bibnamefont {Beyer}}, \bibinfo {author} {\bibfnamefont {J.}~\bibnamefont {G{\"u}termann}}, \bibinfo {author} {\bibfnamefont {C.}~\bibnamefont {Katlein}}, \bibinfo {author} {\bibfnamefont {T.}~\bibnamefont {Krumpen}}, \bibinfo {author} {\bibfnamefont {M.}~\bibnamefont {Bergmann}}, \bibinfo {author} {\bibfnamefont {L.}~\bibnamefont {Hehemann}}, \ and\ \bibinfo {author} {\bibfnamefont {G.}~\bibnamefont {Gerdts}},\ }\href@noop {} {\bibfield  {journal} {\bibinfo  {journal} {Nature Communications}\ }\textbf {\bibinfo {volume} {9}},\ \bibinfo {pages} {1505} (\bibinfo {year} {2018})}\BibitemShut {NoStop}%
\bibitem [{\citenamefont {Browse}\ \emph {et~al.}(2013)\citenamefont {Browse}, \citenamefont {Carslaw}, \citenamefont {Schmidt},\ and\ \citenamefont {Corbett}}]{browse2013impact}%
  \BibitemOpen
  \bibfield  {author} {\bibinfo {author} {\bibfnamefont {J.}~\bibnamefont {Browse}}, \bibinfo {author} {\bibfnamefont {K.}~\bibnamefont {Carslaw}}, \bibinfo {author} {\bibfnamefont {A.}~\bibnamefont {Schmidt}}, \ and\ \bibinfo {author} {\bibfnamefont {J.}~\bibnamefont {Corbett}},\ }\href@noop {} {\bibfield  {journal} {\bibinfo  {journal} {Geophysical research letters}\ }\textbf {\bibinfo {volume} {40}},\ \bibinfo {pages} {4459} (\bibinfo {year} {2013})}\BibitemShut {NoStop}%
\end{thebibliography}
\end{document}